# Microscope Mode Secondary Ion Mass Spectrometry Imaging with a Timepix Detector

Andras Kiss[1], Julia H. Jungmann[1], Donald F. Smith[1], Ron M.A. Heeren[1,a)]

[1] *FOM Institute AMOLF, Science Park 104, 1098 XG Amsterdam, The Netherlands*

In-vacuum active pixel detectors enable high sensitivity, highly parallel time- and space-resolved detection of ions from complex surfaces. For the first time, a Timepix detector assembly was combined with a Secondary Ion Mass Spectrometer for microscope mode SIMS imaging. Time resolved images from various benchmark samples demonstrate the imaging capabilities of the detector system. The main advantages of the active pixel detector are the higher signal-to-noise ratio and parallel acquisition of arrival time and position. Microscope mode SIMS imaging of biomolecules is demonstrated from tissue sections with the Timepix detector.

## I. INTRODUCTION

Atoms, molecules and molecular complexes form the basis of life. They are the building blocks of bio-molecules like peptides, proteins, lipids and DNA. The investigation of the interaction of these macromolecules reveals insights into the dynamic processes that determine the state of biological systems. It is crucial to understand how biological (mal-) function is related to molecular organization. Neurodegenerative diseases, like Alzheimer's, and cancer, for instance, change the cellular biochemistry. Therefore, disease studies and diagnosis can benefit from a fundamental understanding of protein identity, distribution and modification. The investigation of this relationship has given rise to several molecular imaging techniques. In particular, the complexity of biological systems both in the healthy and in the diseased state calls for high spatial resolution molecular imaging techniques that can identify a wide range of biological analytes simultaneously.

Several imaging techniques, such as atomic force microscopy and scanning electron microscopy, enable high spatial resolution analysis of biological systems, but lack chemical identification capabilities. Other techniques, such as fluorescent, antibody or radioactive

---

[a] Author to whom correspondence should be addressed. Electronic mail: heeren@amolf.nl

labeling require prior knowledge about the sample composition and target pre-defined analytes.

Mass spectrometry imaging (MSI)[1, 2] is used to determine the identity and location of many different molecular species from complex surfaces. In particular, this hybrid technique, i.e. mass spectrometry combined with imaging, identifies compounds based on the atomic composition of the sample molecules and their charge state (mass spectrometry) and detects the analytes in a position-correlated way (imaging) without any prior knowledge of the composition of the imaging target. At this point, large-area surface analysis by MSI is used in the several areas of research. In particular, it is applied in the fields of proteomics[3, 4], lipidomics[5-7] and metabolomics[8-10]. Disease studies like the fundamental understanding of the bio-chemistry of neurodegenerative diseases[11, 12] or cancer[13], drug distribution studies[10, 14] and forensics[15, 16], among others, also benefit from the information revealed by MSI.

There are two different approaches for mass spectrometry imaging. Microprobe mode imaging is the most common approach. It uses a highly focused ionization source and every individual pixel of the image is measured separately. Secondary ion mass spectrometry (SIMS) uses a charged primary ion beam for ionization, while matrix-assisted laser desorption/ionization (MALDI) uses a focused laser light source. Many commercial microprobe mode instruments are available for both SIMS and MALDI microprobe MS imaging. It is possible to obtain images with a pixel size below 50 nm with highly focused primary ion beams in SIMS[17, 18] and 5 μm with UV laser probes in MALDI[19-21].

An alternative approach is microscope mode imaging. In this case, the laser or primary ion beam is defocused to irradiate a larger area, typically 100-300 μm in diameter. In this way, microscope mode can reduce analysis time by several orders of magnitude[22]. For example, the analysis of a 100×100 μm area with a 1 μm primary ion beam spot size takes 10,000 individual microprobe mode experiments, while it only takes a single microscope mode experiment that targets an area of 100 μm$^2$. This greatly reduces the measurement time and increases the sample throughput, which is advantageous to the analysis of degrading (biological) samples.

Another advantage of microscope mode MSI in SIMS is that the primary ion gun can be operated with a high primary ion current optimized for higher secondary ion generation. Small primary ion beam spot sizes in microprobe mode (on the order of 500 nm–1 μm) often require relatively low primary ion current (thus low secondary ion count rates), in order to achieve such a spot size and to stay under the static SIMS limit (~$10^{13}$ ions/cm$^2$) and ensure the sample surface is not damaged by the primary ion beam. If the primary ion dose stays below the SIMS static limit, each primary ion samples a fresh position on the surface.

While microprobe mode MSI relies on position-correlated image reconstruction, the microscope mode approach requires that the ion cloud retains its spatial organization in the mass spectrometer and that a position-sensitive detector is used. This means that the obtainable spatial resolution is now decoupled from the ionization source. Thus, the spatial resolution is dependent on the quality of the ion optics in the mass spectrometer, as well as the "magnification factor" of the instrument combined with the capabilities of the position-sensitive detector. The most common detector for this purpose, and the simplest, is the combination of a microchannel plate (MCP) with a phosphor screen and a charged coupled device (CCD) camera behind the screen[22]. Recently, also the direct detection of ions with a specialized CCD camera has been shown[23]. However, these systems cannot simultaneously measure the spatial information and the time-of-flight of different ions. Thus, to spatially map the ion of interest it is necessary to blank out all other ions using an electrostatic blanker. To still obtain the spatial localization of multiple analytes, the sample analysis has to be repeated for each desired analyte. This approach is highly repetitive and hence time-consuming, and additionally not very practical on depleting (biological) samples. On the way towards both position- and time-sensitive measurements, delay-line detectors are another option for microscope mode imaging. These detectors provide both time and space information for all ions simultaneously[24, 25]. However, they lack sufficient multi-hit capability and the image reconstruction is time-consuming. The latter makes the instrument tuning and optimization difficult because of a lack of direct image feedback.

A recent development for microscope mode detection employs in-vacuum pixel detectors, where every pixel acts as an individual detector. The main advantage of these detector systems is the capability to obtain position-[26] and time-resolved[27-29] ion images. The small pixel size of such a detector (here, 55 μm), combined with a high quality ion microscope mass spectrometer with a good magnification factor (typically up to 100×), make these detectors very well-suited for high spatial resolution microscope mode imaging, i.e. a detector pixel (with a physical dimension of 55×55 μm) probes 550 nm on the sample surface at a magnification factor of 100×. Further, such a system provides parallel detection of thousands of ion hits simultaneously. The pixelated detector offers "multiplexed" ion detection capabilities, where each ion hit is detected by multiple pixels, which results in better sensitivity. It also offers direct image acquisition (i.e. no complex instrumental operation or image reconstruction is needed to spatially map all ions from the surface). Furthermore, this class of detectors bears the potential for a fast readout (up to 1 kHz) and hence high experimental repetition rates. The advantages of a pixelated detector have been previously demonstrated using a MALDI ion microscope[26, 28, 29]. However, this technology and its benefits have not yet been applied for SIMS imaging.

Secondary ion mass spectrometry was the first mass spectrometry imaging approach. It utilizes high energy charged particles to generate secondary ions from a surface. These high energy particles are usually elemental ions from a liquid metal ion source (LMIS). Further, the primary ion beam can easily be focused to micrometer and nanometer probe sizes. Compared to MALDI, SIMS suffers from extensive fragmentation of secondary ions and from low secondary ion yield. Also, ion counts steeply drop at higher masses. This limits its use mostly to the analysis of low mass molecular and elemental ions. Recent developments in sample preparation methods[30-32] and the introduction of cluster primary ion sources ($Au_n^{m+}$, $Bi_n^{m+}$, $C_{60}^+$)[33-35] have improved the secondary ion yields for lipids and small peptides. These developments have helped move the technique from applications in the domains of surface physics and solid state physics to the study of biological samples such as cells and tissue sections, where it proved to have an outstanding spatial resolution of few hundred nanometers compared to other mass spectrometry imaging methods.

In this work, we present the first example of an in-vacuum pixelated detector for microscope mode SIMS imaging. The detector used was a member of the Medipix/Timepix family in combination with a chevron MCP stack. The advantages of the Timepix pixelated detector for microscope mode MSI, i.e. position- and time-sensitive measurements, measurement of different masses simultaneously, high dynamic range, high throughput and high spatial resolution, are transferred to SIMS microscope mode MSI. The detector was evaluated for SIMS imaging with a benchmark sample, as well as biological tissue sections. The different acquisition modes of the Timepix active pixel detector, i.e. the particle counting mode and the time-over-threshold (both in microprobe mode) and the time-of-flight modes (in microscope mode), are evaluated for SIMS imaging. Whenever applicable, the performance of the new detection system is compared to that of an established detector system.

## II. EXPERIMENTAL

### A. The Medipix/ Timepix Detectors

The detectors in the Medipix/Timepix family are based on complementary metal-oxide-semiconductor (CMOS) technology. The detectors have been developed by the Medipix collaboration hosted by CERN[36]. A single chip contains 256×256 pixels (pixel size=55 µm×55 µm) and it is possible to build chip arrays of 2×2n (n=1,2…) chips.

The Timepix chip[37] is the successor of the Medipix2 chip[38], where each pixel contains an additional clock with a maximum clock speed of 100 MHz and measurement interval of 118 µs. This makes it possible to use the Timepix chip in three different operating modes. The first is Medipix counting mode, which simply counts the number of particles that hit the individual pixels of the chip in a set time frame ("particle counting"). The second is called time-over-threshold (TOT) mode where the time that a pixel spends above a certain charge threshold ("intensity distribution" of how much charge is deposited per pixel) is measured. Finally, the Timepix has a time-of-flight (TOF) mode that is capable of measuring the arrival time of particles with respect to an external trigger signal. This mode is particularly well-suited for time-of-flight mass spectrometry (TOF-MS), since every pixel functions as an individual TOF detector.

The data is read out by a 1 Gbit/s read-out system developed by the ReLAXD project (high Resolution Large Area X-ray Detector[39, 40]) for 2×2 chip assemblies. It reads out the four chips in parallel, which results in a maximum frame rate of 30 frames/s. The chips are controlled by a dedicated acquisition and control software package, Pixelman[41, 42].

### B. Imaging setup on the ion microscope

All experiments were performed on a Triple Focusing Time of Flight mass spectrometer (TRIFT-II, Physical Electronics, INC., Chanhassen, USA) equipped with a gold Liquid Metal Ion Gun (LMIG) primary ion source. All experiments were performed in the positive-ion mode with 22 keV $Au^+$ primary ions. The ion source routinely operates with a repetition rate in the kHz range. Thus, to synchronize the mass spectrometer with the Timepix detector (in TOF mode), the original trigger signal from the instrument needed to be down-sampled to 30 Hz (the maximum readout rate of this setup). The Timepix detector can only collect one measurement frame before it must be readout (i.e. it is single-stop) and cannot measure during the readout phase. Thus, an experimental repetition rate higher than the maximum readout rate would not result in any additional measurements on the Timepix. For the evaluation of the particle counting and TOT mode (sections III.A. and III.B.), the mass spectrometer is operated in "quasi" microprobe mode, i.e. the primary ion source is rastered with a 100-200 µm wide square tile with a raster speed in the MHz range, where every tile measures 256×256 pixels. However, the images are directly acquired with the position-sensitive Timepix detection system and need not be reconstructed by position-correlation, as is required with traditional microscope mode experiments. Note that for these experiments, the Timepix detection system and the TRIFT instrument run completely asynchronously. The primary ion source runs at a repetition rate of 8 kHz, while the Timepix detector collects data at 30 Hz. Hence, the Timepix collects ions from multiple primary ion gun raster postions in one acquisition frame.

For time-resolved measurements (evaluation of the TOF mode, sections III.C-III.E), the rastering was disabled and the ion beam was defocused for microscope mode operation. In this way, the imaged area was about 100 µm wide and 300 µm high.

The detection system was a chip assembly consisting of bare Timepix ASIC (application-specific integrated circuit) in a 2×2 array mounted 2 mm behind a chevron MCP stack ($\phi$ = 4 cm, 12 µm pores, 15 µm pitch). Details of the detector assembly can be found elsewhere[28]. The MCP was used at a bias of 1.9 kV (gain: $5 \cdot 10^6$) unless stated otherwise. The potential between the backside of the MCP and the chip was 600 V.

C. Reference measurement system

The high voltage line from the MCP backside was decoupled into the original electronics rack of the TRIFT-II mass spectrometer, which consists of a constant-fraction discriminator (CFD) and a multi-stop time-to-digital converter (TDC) with 138 ps time bins. This setup makes it possible to acquire reference spectra with an established method simultaneously with the Timepix measurements. Reference images are acquired by position-correlated registration of the ion current, i.e. traditional microprobe mode.

**D. Timepix-generated spectra and images**

The Timepix detector was operated in all of its possible operating modes. In all modes, a typical, sparse data frame contains the x- and y-coordinate of every triggered pixel. Additionally, per acquisition mode, the data file lists the number of particles counted by the pixel (counting mode), the time during which the pixel was over threshold (TOT mode) or the time-of-flight (TOF mode). The Medipix counting and the TOT mode are imaging modes (used in combination with the microprobe mode). Images are integrated by adding the Medipix or TOT counts of several image frames, respectively. In the TOF mode, every measurement frame contains the TOF information obtained from a single primary ion pulse. The mass spectrum is reconstructed by making a histogram of the TOF values from the separate frames. Standards were used to calculate mass calibration parameters, which were then applied to the TOF spectra. Total ion images were constructed by summing all of the individual frames. Selected ion images were plotted by extracting the pixel positions for a selected mass spectral peak.

**E. Samples**

An organic dye (green Staedtler Lumocolor 318-5 permanent marker, Staedtler Mars GmbH & Co. KG, Nuernberg, Germany) and a solution of polyethylene glycol (2 mg/ml PEG; 200-3500 dissolved in methanol) spotted on an indium tin oxide (ITO) coated glass slide (4-8 Ώ resistance, Delta Technologies, Stillwater, MN) were used for initial evaluation of the new experimental setup. A hexagonal thin bar transmission electron microscopy (TEM) grid (700 mesh, G2760N, 3.05 mm diameter, 37 μm pitch, 8 μm bar width; Agar Scientific Limited, Stansted, United Kingdom) was placed on top of the wet solutions, which allowed the standards to be detected from underneath the grid.

Mouse testis tissue (male balb/c mouse; Harlan Laboratories, Boxmeer, The Netherlands) was sectioned into 12 μm thin sections using a Microm HM525 cryomicrotome (Thermo Fisher Scientific, Walldorf, Germany) and then thaw-mounted on an ITO slide. After sectioning, the tissue sections were kept at -20 °C until further use. The sample was dried in a vacuum desiccator for 30 minutes and then coated with 1 nm gold layer using a sputter coater equipped with a FT7607 quartz crystal microbalance stage and a FT690 film thickness monitor (Quorum Technologies SC7640, New Haven, East Sussex, United Kingdom) prior to measurement

**III. RESULTS**

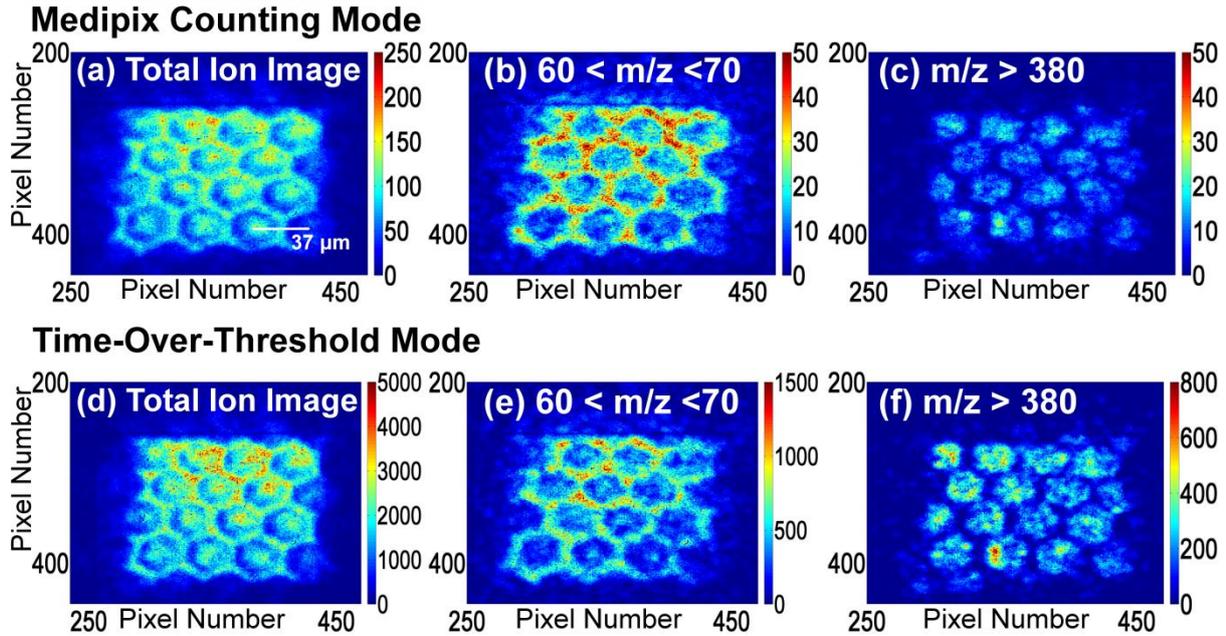

FIG. 1. Comparison of Medipix counting mode (a-c) and Time-Over-Threshold mode (d-f) of the Timepix detector using the total ion image (a,d) and the selected ion images of copper (b,e) and the organic dye (c,f).

**A. Ion imaging: Medipix versus time-over-threshold mode acquisitions**

Experiments were performed to assess the image quality of the Timepix. Therefore, the TEM grid/organic dye sample was measured in both ion counting mode and TOT mode ("intensity mode"). The results of these experiments are shown in Figure 1.

Both modes return TEM grid images with a very good image quality and contrast (Figure 1). A comparison of the images of the two modes reveals that the contrast in TOT mode is better than in the counting mode. In TOT mode, the time each pixel spends above a certain signal threshold is measured which results in a better signal-to-noise ratio (S/N) and better image contrast than in the counting mode.

When the Timepix is run in the Medipix counting or the TOT imaging mode, mass-selected images can only be created by selectively blanking all unwanted ions by the use of an electrostatic blanker[22] (unlike in the Timepix TOF mode, sections III.C-III.E). As stated earlier, mass-selection by an electrostatic blanker is suboptimal. However, it can readily be employed for the evaluation of the image quality delivered by the Timepix system. The motivation is to assess if the image quality (image resolution) of the Timepix is acceptable or better than current approaches for SIMS imaging. Thus, later experiments could include "dual" or even "triple" mode experiments, in which different pixels of the detection system are programmed to operate in the counting, the TOT or the TOF mode (which will again remove the need for mass-selection with the blanker).

As expected, the selection of the copper ion mass range ($60<m/z<70$) yields a high contrast and high resolution image of the grid bars (Fig. 1, b,e), while a selection of ions above $m/z=380$ results in a complementary ion image of green dye in the holes (Fig. 1 c,f).

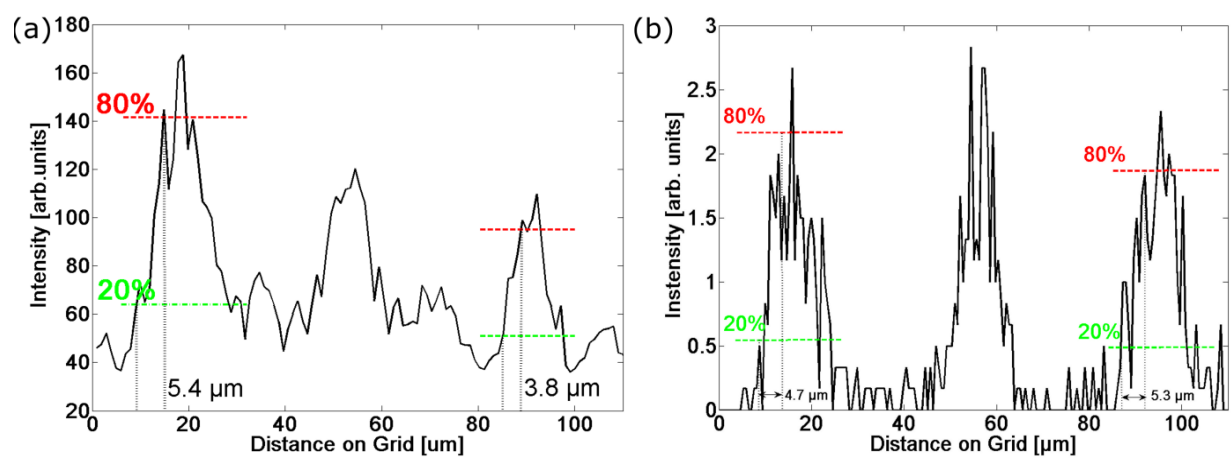

FIG. 2. Line scans from the copper signal of a TEM grid used to calculate the spatial resolving power. Acquired with the Timepix detector (a) and the original MCP and TDC detector system (b).

## B. Spatial resolving power: Timepix versus standard TRIFT images

The spatial resolving power, rather than the spatial resolution (1 detector pixel corresponds to 550 nm-1μm on the sample surface depending on the ion optical magnification factor in the TRIFT-II MS), of the images shown in Figure 1 is investigated and compared to the spatial resolving power that is obtained in the equivalent "postion-correlated" microprobe experiment. Generally, the spatial resolving power is determined by generating an intensity profile across features of the sample, a so-called line scan, and assessing the distance between the 20% and 80% points of the peaks in the line scan[43]. Figure 2a shows the result of such a line scan through Fig. 1b (Medipix counting mode, copper ions, *m/z*=60-70), which yields a spatial resolving power of ~5 µm (Figure 2a). The spatial resolving power of the Timepix detector in TOT mode and the spatial resolution from Timepix TOF microscope mode imaging is similar[44]. The spatial resolving power of the position-correlated microprobe mode image reconstruction is similar but slightly inferior to the Timepix resolution (Fig. 2b)[44]. Thus, the spatial resolution of the two detection methods, position-sensitive detection with the Timepix and position-correlated image reconstruction on the basis of the MCP signal, is comparable with a slight performance advantage for the Timepix. Also note that the absolute intensity of the signal obtained with the Timepix detector is much higher than the one measured with the MCP only. This is due to the multiplexed detection capabilities of the Timepix detector where a single ion hit is translated into an electron shower by the MCP which is then detected on multiple pixels on the Timepix detector. It is also worth noting that the Timepix image was measured using an MCP bias of 1.3 kV (gain of $2 \cdot 10^5$), while the standard MCP-based image detection required a bias of 1.75 kV (gain of $1.5 \cdot 10^6$) for any ion to be detected. A lower MCP bias (and hence gain) results in a longer MCP lifetime.

## C. Mass Spectra: Timepix versus TRIFT TDC

The previous sections evaluate the Timepix imaging quality in Medipix counting mode and TOT mode. These modes allow fast image acquisition of total ion images (see section III.E for an evaluation of the acquisition speed and obtainable repetition rates). Thus, high S/N images can be collected in a short amount of time. However, the measurement of selected ion images for every mass spectral peak are still time-consuming (i.e. only a single mass can be imaged at once) in these two imaging modes. Thus, the desired Timepix mode of operation for TOF-SIMS imaging is the TOF mode.

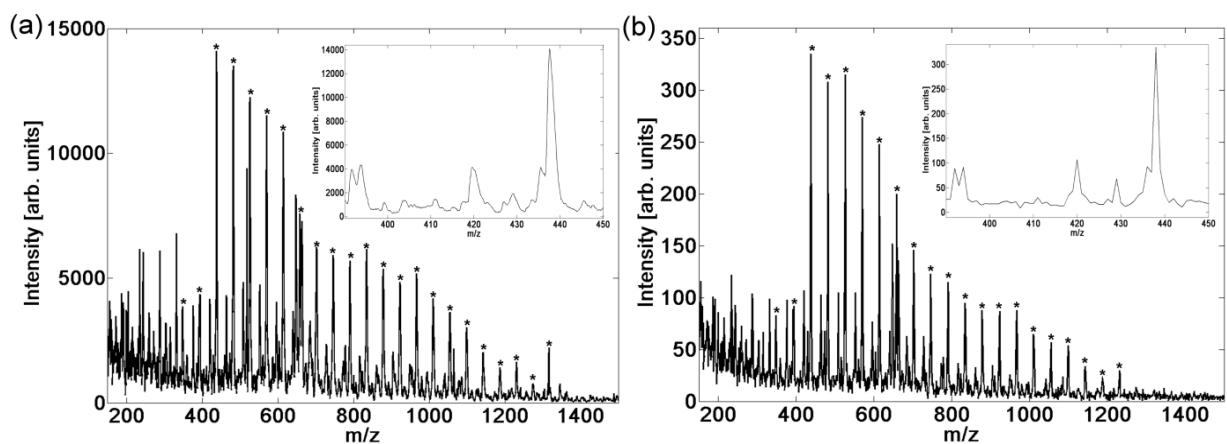

FIG. 3. Mass spectrum of PEG mixture measured with the Timepix detector (a) and with the MCP-TDC (b). The sodiated ions of different length PEG oligomers are marked (*). Insets show a selected peak at *m/z* 437 ([PEG$_9$+Na]$^+$ ion).

The quality of Timepix-acquired SIMS mass spectra is evaluated with a mixture of PEG. For this measurement, the MCP was biased at a voltage of 2.3 kV (gain of $10^7$) and the repetition rate of the SIMS source was down-sampled to 30 Hz. This high gain was needed in order to record a mass spectrum with the reference TDC detector system. The comparison of the two detectors shows that all peaks detected with the TDC are also present in the Timepix spectrum (Figure 3). The detected peaks are the sodiated ions of different length PEG oligomers. Both systems achieve the same mass range. The signal-to-noise ratio of the mass spectrum acquired with the Timepix detector (peak at *m/z*=437 detected with a S/N of 18) is almost 40% superior to the one measured with the TDC setup (peak at *m/z*=437 detected with a S/N of 13). The Timepix spectral mass resolution is slightly less, due to the relatively long clock cycles of the detector (10 ns, detailed evaluation in reference 26). Since the ion mass is determined from its time-of-flight in TOF-SIMS measurements, the minimum clock cycle length determines the obtainable mass resolution. A clock cycle of 10 ns corresponds to approximately 150 mDa for an ion around 400 Da on our ion microscope mass spectrometer

**D. Time- and space-resolved microscope mode SIMS imaging**

Figure 4 illustrates time-resolved (i.e. mass-selected) imaging of the benchmark sample (green organic dye under a TEM grid) using the TOF mode of the Timepix detector. The images of the low mass copper ions show good contrast (Figure 4a, red). The images of the organic dye (in the holes of the grid, Figure 4b, green) are relatively sparse due to low ion counts. These results show that the low ion count rates encountered with SIMS result in a low S/N, especially for higher mass ions, with the Timepix detector. The low ion count rates encountered here are the result of multiple factors, the most important of which is the low repetition rate of the gold primary ion source. It was necessary to decrease the repetition rate of the primary ion source by a factor of 266 (from 8 kHz to 30 Hz), in order to synchronize it with the Timepix detector. Since there is a linear connection between the repetition rate of the ion source and the number of ions delivered to the sample, the latter decrease by the same factor. Similarly, the aquisition time must be increased by the same factor to reach the same primary ion dose on the surface as in a microprobe experiment. Another factor that causes low ion counts is the lack of a post-acceleration voltage in this Timepix setup where the Timepix detection system is held at ground potential  A post-acceleration voltage is particuallarly important for higher mass ions, since they impinge on the detection system with a lower momentum and hence might not be able to initiate an electron cascade in the MCP. Combined, these effects explain the observed steep drop in secondary ion counts with increasing mass. The lower ion yield (per frame) of the SIMS setup versus a MALDI experiment is clearly visible if we compare the total ion image from a single measurement frame from both SIMS and MALDI[44].

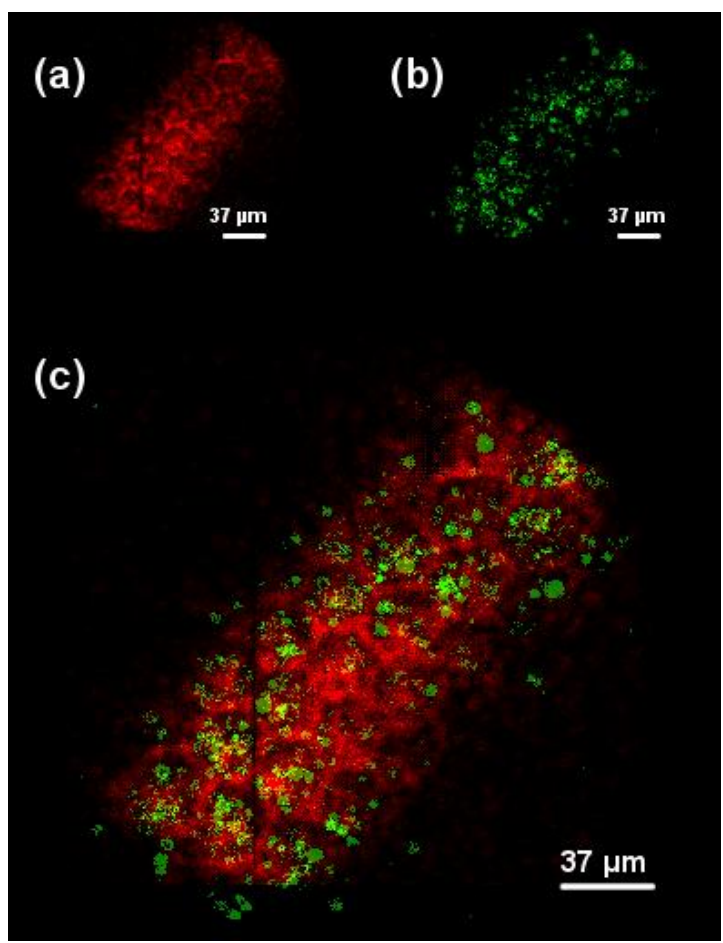

Figure 4. Time-resolved overlay image of the green organic dye standard below a TEM grid. The plotted masses are *m/z*=63 (copper grid, a), *m/z*=385 (organic dye, b) and the overlay of the two images (c).

**E. SIMS-MSI on biological samples with the Timepix**

Figure 5 demonstrates successful microscope mode SIMS imaging of a biological tissue section, mouse testis, with the Timepix detector in TOF mode. Here, the detector and the primary ion source are synchronized and more than 300,000 frames were collected at this one position (more than 3 h measurement at 30 Hz) due to the 30 Hz experimental repetition rate of the Timepix system. It is possible to distinguish different features in the measurement area. The selected ion images show ions present on the tissue section, such as cholesterol (*m/z*=369, green) and another organic ion at *m/z*=358 (blue). For contrast, indium (*m/z*=115, red) was selected which was detected mostly from the surface of the ITO-coated glass slide. The diagonal "line" across the image marks the tissue boundary.

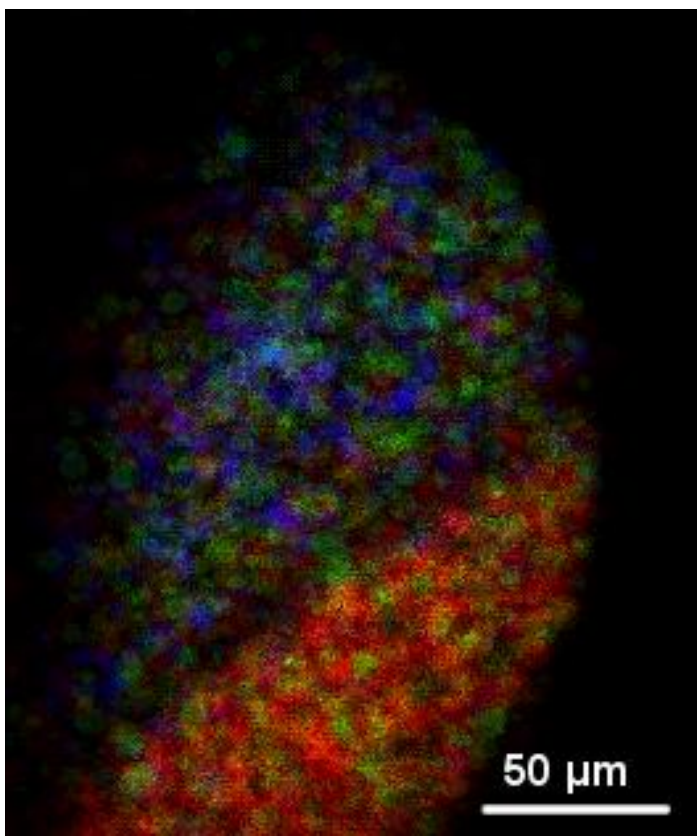

FIG. 5. Overlay of different time resolved images of a mouse testis section. The selected ion images are *m/z* 115 (Indium, red), *m/z* 369 (cholesterol-OH, green) and *m/z* 358 (blue).

## IV. CONCLUSION

For the first time, a MCP/Timepix detector assembly has been combined with a SIMS mass spectrometer for microscope mode SIMS imaging. This combination enables high spatial resolution imaging of biomolecules from tissues with high sensitivity due to the parallel detection of ions achieved with the pixelated detector. The detector system's performance was compared to the established detection system of the SIMS instrument. The signal-to-noise ratio of the spectra acquired with the Timepix detector proved to be significantly better at a comparable spectral mass resolution. Also, the spatial resolving power of the Timepix-generated images is superior to the original detection system. Additionally, it is advantageous that the MCP in the detector assembly can be run at sub-saturation gains without compromising the measurement quality, which is advantageous for the lifetime of the MCP and enables tuning the MCP gain to achieve optimal spatial resolution.

This paper shows high quality ion images from biological surfaces both in ion counting and time-of-flight mode with a Timepix detector. It also shows the capability of the Timepix to easily deliver time-resolved ion images from the analysis of a tissue section using the time-of-flight mode of this active pixel detector.

The biggest challenge for the SIMS-Timepix setup is low ion counts per frame. Normally, the relatively low SIMS ion load is compensated by the use of a high repetition rate primary ion source, which allows the summation of many scans per pixel. This is currently not feasible in combination with a Timepix detection system since the data readout of the Timepix system is limited to tens of frames per second, which increases the overall measurement time. This necessitates a new version of Timepix readout electronics with a duty cycle in the kHz range. Furthermore, the use of a primary ion source with higher secondary ion yield, such as the Au LMIG in cluster mode or a $C_{60}$ ion source should alleviate the low secondary ion count rates. Previous studies show that the use of $C_{60}$ as a primary ion improves secondary ion yield (versus monatomic primary ions) between a factor of 13×-300×[34, 45-49]. With a mild 50-fold improvement in secondary ion yield, the time needed for the experiment shown in Fig. 5 would be reduced from ~3 hours to only 3.6 minutes. Larger primary ion species also provide better ionization of larger molecules (and less fragmentation). Thus, the combination of such a primary ion source with the Timepix detector for microscope mode imaging is very attractive.

**ACKNOWLEDGEMENTS**

This work is part of the research program of the Foundation for Fundamental Research on Matter (FOM), which is part of the Netherlands Organisation for Scientific Research (NWO). Part of this research is supported by the Dutch Technology Foundation STW, which is the applied science division of NWO, and the Technology Program of the Ministry of Economic Affairs, Project OTP 11956. This publication was supported by the Dutch national program COMMIT and the Netherlands Proteomics Center. The authors acknowledge Sheng Chen (Physical Electronics) for excellent support during this project by sharing his knowledge on the TRIFT II system. The authors are thankful to Frans Giskes and Ronald Buijs for their technical support.

**Supplementary figures:**

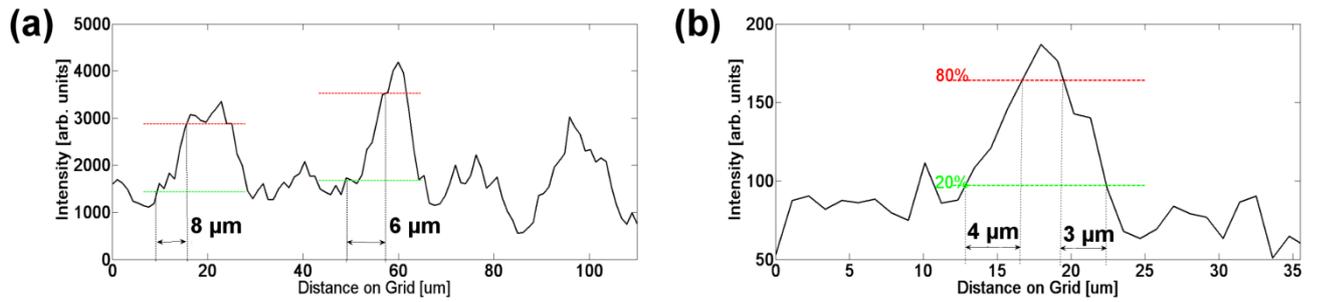

FIG. S1. Line scans from the copper signal of a TEM grid used to calculate the spatial resolving power. Acquired with the Timepix detector in Time-over-Threshold mode (a) and in Timepix TOF mode (b).

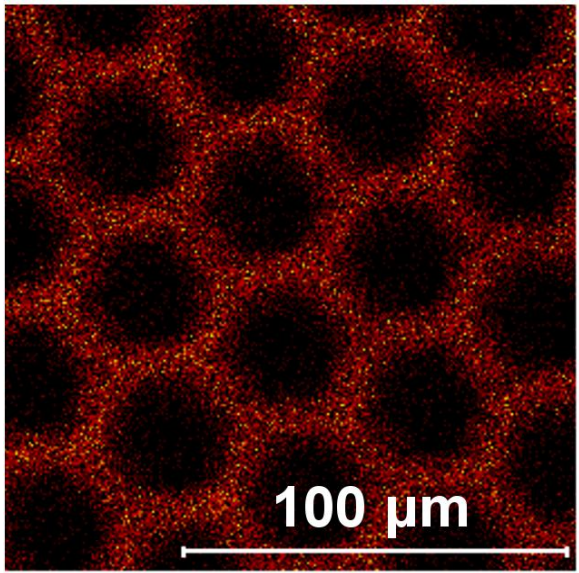

FIG. S2. Image of the TEM grid acquired with the MCP-TDC detector system

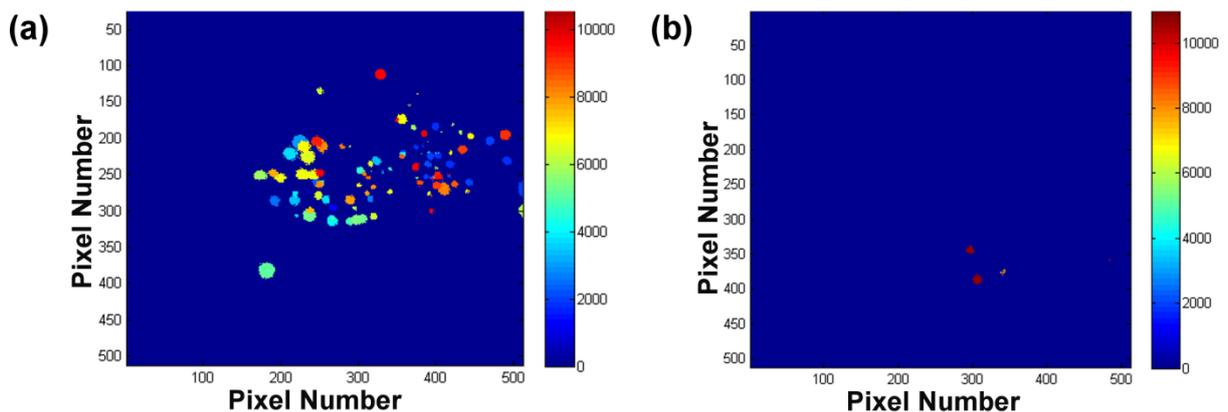

FIG. S3. Comparison of a single frame of ion events detected on the Timepix detector using MALDI (a) and SIMS (b). Per frame, SIMS has a much lower number of ions, which necessitates many scans to be collected for sufficient signal-to-noise ratio.

18